\documentclass[]{aa}
\usepackage{epsfig}

\begin{document}
\title{Discovery of a dust cloud next to $\sigma$\,Orionis}
\author{Jacco Th. van Loon and Joana M. Oliveira}
\institute{Astrophysics Group, School of Chemistry \& Physics, Keele
           University, Staffordshire ST5 5BG, United Kingdom}
%\offprints{\email{jacco@astro.keele.ac.uk}}
\date{Received date; accepted date}
\titlerunning{Discovery of a dust cloud next to $\sigma$\,Orionis}
\authorrunning{van Loon \& Oliveira}
\abstract{
We report on the discovery of a mid-infrared source at a projected distance of
only 1200 AU from the O9.5\,V star $\sigma$\,Orionis. The spatially resolved,
fan-shaped morphology and the presence of an ionization front, as well as
evidence in the spectrum for processed dust grains, all suggest that it is a
proto-planetary disk being dispersed by the intense ultraviolet radiation from
$\sigma$\,Orionis. We compute the mass budget and the photo-evaporation
timescale, and discuss the possible nature of this remarkable object.
}
\maketitle

\section{Introduction}

\subsection{Proto-planetary disks}

The conservation of angular momentum in a collapsing and fragmenting molecular
cloud naturally leads to the presence of circumstellar disks around young
stellar objects. It is believed that by way of co-agulation of dust and
subsequent accretion of gas planets may form in these disks --- whence their
name ``proto-planetary disk''.

The evolution of a proto-planetary disk into a planetary system must compete
with a multitude of disk-dispersal mechanisms: accretion, stellar wind and
photo-evaporation by the central star depletes the disk from the inside,
whilst nearby hot, massive O- and B-type stars photo-evaporate the disk from
the outside. Censuses of circumstellar disks in several young stellar
clusters, though prone to observational bias, suggest that most disks
disappear in $\sim6$ Myr (Haisch, Lada \& Lada 2001).

Photo-evaporation of disks by external stars has been observed to happen in
the Orion Nebula (O'Dell, Wen \& Hu 1993) and in the massive star forming
region NGC\,3603 (Brandner et al.\ 2000). In these harsh environments, the
effect of the intense UV radiation is expected to disperse the disks on
timescales of only $\sim10^5$ yr, which is much quicker than the ages of the
young clusters in the Orion Nebula (1--2 Myr: McCaughrean \& Stauffer 1994)
and NGC\,3603 (3 Myr: Hofmann, Seggewiss \& Weigelt 1995).

\subsection{$\sigma$\,Orionis and the $\sigma$\,Orionis cluster}

At a Hipparcos distance of 352 pc, the bright ($m_{\rm V}=3.8$ mag) O9.5\,V
star $\sigma$\,Orionis is the primary component of a quintuple system of O and
B stars (ADS\,4241). ADS\,4241\,AB sits at the heart of the most massive
visual binary known, with estimated masses of $M_{\rm A}\simeq25$ M$_\odot$
and $M_{\rm B}\simeq15$ M$_\odot$, an orbital period $P=158$ yr, semi-major
axis $a=0.265^{\prime\prime}$, eccentricity $e=0.06$ and inclination
$i=153^\circ$ (Heintz 1997). The early B-type companions C and D are much
farther away, at projected distances of $d\simeq11^{\prime\prime}$ to the West
and $d\simeq13^{\prime\prime}$ to the East, respectively. At
$d\simeq42^{\prime\prime}$ to the East-Northeast, the helium-rich B2\,Vp star
$\sigma$\,Ori\,E is the most remote companion.

This massive multiple system is part of the Orion OB1 association (Brown, de
Geus \& de Zeeuw 1994) and forms the core of a recently discovered (Walter,
Wolk \& Sherry 1997) young cluster of stars, brown dwarfs and isolated
planetary mass objects (Zapatero Osorio et al.\ 2000), with an age of $\sim4$
Myr (Oliveira et al.\ 2002). Its immediate vicinity is largely free of dust
and molecular material. This cluster is in a crucial phase in terms of disk
dispersal, and we have obtained IR data (L- and N-band) to detect and
characterize circumstellar disks around cluster members across the entire mass
range (Oliveira et al.\ 2003).

One of the peculiarities of $\sigma$\,Orionis that so far has received little
attention is its association with a bright source of mid-IR emission,
IRAS\,05362$-$0237. One might try to attribute it to free-free emission from
the boundary between the colliding winds of ADS\,4241\,A and B, but this
interpretation is not supported by the shape of the IR spectral energy
distribution, and its X-ray luminosity is typical for the relatively weak
stellar wind of $\sigma$\,Orionis with a mass-loss rate of $\dot{M}<10^{-8}$
M$_\odot$ yr$^{-1}$ (Chlebowski \& Garmany 1991). Nevertheless, when radio
emission was detected from a position 2--3$^{\prime\prime}$ to the North of
$\sigma$\,Orionis it was again attributed to (probably non-thermal) emission
from ADS\,4241\,AB, with the positional mismatch between the accurate radio
position and the USNO optical position of $\sigma$\,Orionis deemed
insignificant (Drake 1990).

\section{Sub-arcsecond mid-IR observations of $\sigma$\,Orionis}

The mid-IR imager and spectrograph TIMMI-2 at the ESO 3.6m telescope at La
Silla, Chile, was used on the night of 15/16 December, 2002. Images of
$\sigma$\,Orionis were obtained through the N1-band filter ($\lambda_0=8.6$
$\mu$m, $\Delta\lambda=1.2$ $\mu$m for Full-Width at Half Maximum (FWHM) and
$\Delta\lambda=1.7$ $\mu$m between blue and red cut-off) and through the
Q1-band filter ($\lambda_0=17.75$ $\mu$m, $\Delta\lambda=0.8$ $\mu$m for FWHM
and $\Delta\lambda=1.4$ $\mu$m between blue and red cut-off). The pixel scale
was $0.2^{\prime\prime}$ pixel$^{-1}$, resulting in a
$64^{\prime\prime}\times48^{\prime\prime}$ (RA$\times$Dec) field-of-view. We
used a chop throw of $10^{\prime\prime}$ in the N-S direction and a nod offset
of $10^{\prime\prime}$ in the E-W direction. The resulting stellar images had
a FWHM of 0.7--0.8$^{\prime\prime}$. Photometry was performed on the
shift-added images, using a circular software aperture with a
$2^{\prime\prime}$ diameter, and calibrated against HD\,4128 and HD\,32887.

TIMMI-2 was used on the same night for spectroscopy. With a slit of
$3^{\prime\prime}$ wide and $50^{\prime\prime}$ long, the spectral resolving
power, limited by the pixel scale of 0.02 $\mu$m pixel$^{-1}$, was
$R\sim200$--300 across a useful window of $\lambda=8$ to 13 $\mu$m. The
spectrum was flux-calibrated against HD\,4128 and HD\,32887 as well as the
N1-band photometry.

\section{Discovery of $\sigma$\,Ori IRS1}

\subsection{Imagery}

The discovery of {\em another} bright mid-IR source (Fig.\ 1) --- which we
designate $\sigma$\,Ori IRS1 --- next to $\sigma$\,Orionis came totally
unexpected. The new object has a compact core at only $3.3^{\prime\prime}$ to
the North-Northeast from $\sigma$\,Orionis, and also exhibits extended
emission in a fan-shaped morphology, pointing away from $\sigma$\,Orionis. The
image in Fig.\ 1 was obtained after one iteration of a deconvolution algorithm
(Lucy 1974) within the ESO software package {\sc midas}, using
$\sigma$\,Orionis as a model for the Point Spread Function.

%
% FIGURE 1
%
\begin{figure}[tb]
\centerline{\psfig{figure=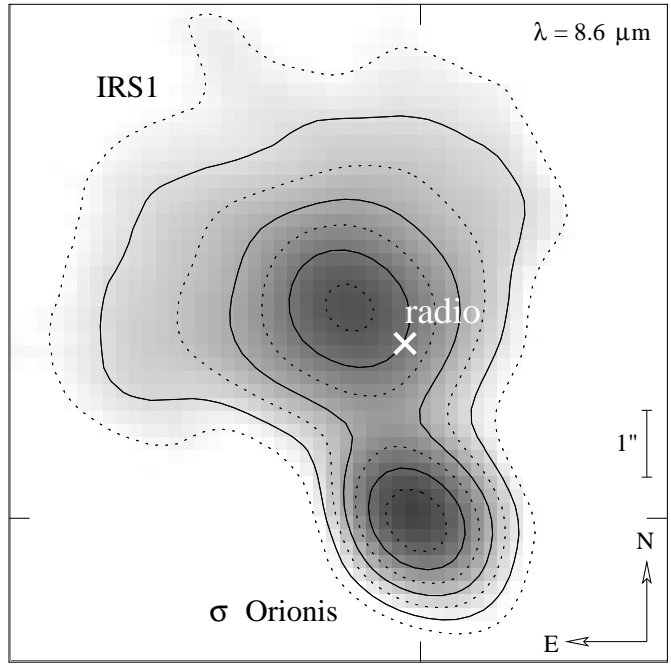,width=88mm}}
\caption[]{Mid-infrared image of $\sigma$ Orionis (near the bottom) and IRS1
(near the centre), taken through the N1-band ($\lambda_0=8.6$ $\mu$m) filter
with the TIMMI-2 camera at the ESO 3.6m telescope. The position of the radio
source as detected by Drake (1990) is indicated by a white cross. The field
measures $10^{\prime\prime}\times10^{\prime\prime}$, with North up and East to
the left, and is displayed on a logarithmic scale with a dynamic range of a
factor 100.}
\end{figure}

The close proximity of IRS1 to the mag 4 star $\sigma$\,Orionis is the main
reason why this object has so far eluded discovery, and it is only thanks to
the greatly reduced brightness contrast in the mid-IR that it was found. In
fact, the core of IRS1 is with an N1-band flux density of $N_{\rm
core}=0.573\pm0.029$ Jy (circular aperture of $2^{\prime\prime}$ diameter)
nearly equally bright as $\sigma$\,Orionis, which has $N=0.617\pm0.031$ Jy.
The core of IRS1 is marginally resolved in the N1-band: its FWHM is
$1.1^{\prime\prime}$, about $1.5$ times greater than the FWHM of
$\sigma$\,Orionis.

To estimate the contribution of the extended emission to the total integrated
emission from IRS1, we also measured the brightness of IRS1 within a
rectangular box of $4.4^{\prime\prime}\times3.2^{\prime\prime}$ (comparing
with the brightness of the standard stars within an identical area), and found
$N_{\rm core+extended}=0.901\pm0.045$ Jy. The IRAS 12 $\mu$m flux density is
$S_{12}=4.5\pm0.2$ Jy\footnote{The IRAS flux densities were redetermined from
the original scans on the IRAS data server in Groningen.}, the Q1-band
brightness of the core of IRS1 is $Q_{\rm core}=2.38\pm0.24$ Jy, and the IRAS
25 $\mu$m flux density is $S_{25}=15\pm2$ Jy.

\subsection{Spectroscopy}

%
% FIGURE 2
%
\begin{figure}[tb]
\centerline{\psfig{figure=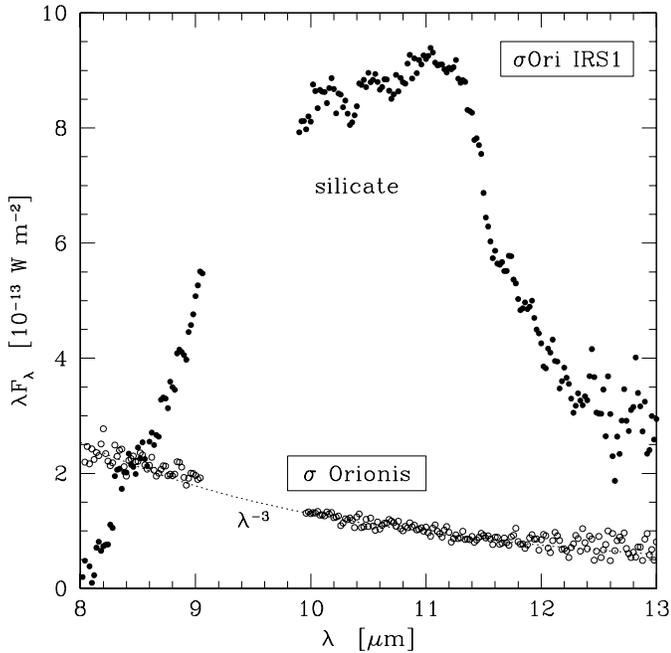,width=88mm}}
\caption[]{N-band spectra of $\sigma$ Orionis (circles) and $\sigma$\,Ori IRS1
(dots). The spectrum of $\sigma$ Orionis shows a pure photospheric continuum,
but the spectrum of $\sigma$\,Ori IRS1 exhibits a strong, broad emission
feature indicative of silicate dust.}
\end{figure}

N-band spectra were obtained {\em simultaneously} for $\sigma$\,Orionis and
the core of $\sigma$\,Ori IRS1 (Fig.\ 2). The spectrum of $\sigma$\,Orionis
follows the Rayleigh-Jeans tail of the photospheric emission of this hot star,
$\lambda F_\lambda \propto \lambda^{-3}$. The spectrum of $\sigma$\,Ori IRS1,
however, is dominated by a strong and broad emission feature, which we
attribute to silicate dust. The silicate feature has weak shoulders at
$\lambda\simeq8.6$ and 11.7 $\mu$m and, although the spectral region between
$\lambda=9$ and 9.9 $\mu$m was useless due to a defunct channel in the TIMMI-2
array, it appears to peak around $\lambda\simeq11$ to 11.3 $\mu$m. The
underlying continuum is very red, with a flux ratio between $\lambda=13$ and 8
$\mu$m of $F_{13}/F_8\simeq10$.

\section{Analysis}

\subsection{Gas properties}

The fan-shaped morphology of the extended mid-IR emission from $\sigma$\,Ori
IRS1 suggests that it is acted upon by the intense radiation field of $\sigma$
Orionis, either through radiation pressure on dust grains or through
photo-evaporation of gas interspersed with the dust. The projected distance
beween $\sigma$\,Ori IRS1 and $\sigma$\,Orionis is only $d\simeq1200$ AU! The
stellar wind of the main-sequence star $\sigma$\,Orionis, however, is too weak
to have a significant impact.

The radio emission is consistent with optically thin free-free emission at
wavelengths of $\lambda=2$ and 6 cm, possibly becoming optically thick around
a wavelength of $\lambda\simeq15$ cm (Fig.\ 3). The latter can be used to
estimate the electron density in the emission region (Osterbrock 1974):
$\tau=8.24\times10^{-2}T_{\rm e}^{-1.35}\nu^{-2.1}\int{n_+n_{\rm e}{\rm d}s}$.
For $\tau=1$, with an electron temperature $T_{\rm e}=10,000$ K, frequency
$\nu=2$ GHz and pathlength ${\rm d}s\simeq10^{-3}$ pc, the electron density is
estimated to be $n_{\rm e}=n_+\simeq10^6$ cm$^{-3}$, which is indicative of a
relatively dense, ionized region. The radio position (Drake 1990) is located
$2.63^{\prime\prime}$ North and $0.18^{\prime\prime}$ East from
$\sigma$\,Orionis, at the rim of the extended mid-IR emission facing
$\sigma$\,Orionis. This suggests that the free-free emission arises from a
photo-ionized region at the interface between the radiation field of
$\sigma$\,Orionis and the dust region in $\sigma$\,Ori IRS1.

If the ionizing radiation from $\sigma$\,Orionis is absorbed in a layer with a
thickness comparable to the radius $r$ of the ionization front, then,
following Bally \& Reipurth (2001), the electron density is $n_{\rm
e}\simeq(L_{\rm H}/4\pi\alpha_{\rm B}r)^\frac{1}{2}d^{-1}$. With a Lyman
continuum photon rate from $\sigma$\,Orionis of $L_{\rm H}\simeq10^{48}$
s$^{-1}$ (Peimbert, Rayo \& Torres-Peimbert 1975), recombination coefficient
$\alpha_{\rm B}=2.6\times10^{-13}$ cm$^3$ s$^{-1}$ (Osterbrock 1974), distance
$d\simeq1200$ AU from $\sigma$\,Orionis and size $r\simeq300$ AU, we obtain
$n_{\rm e}\simeq0.5\times10^6$ cm$^{-3}$. This is consistent with the value
estimated from the radio emission, which implies that IRS1 cannot be much
further away from $\sigma$\,Orionis than the projected $d\simeq1200$ AU.

%
% FIGURE 3
%
\begin{figure}[tb]
\centerline{\psfig{figure=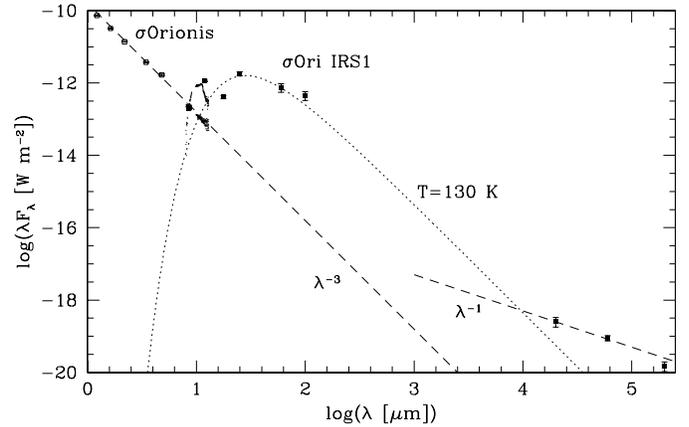,width=88mm}}
\caption[]{Spectral energy distributions of $\sigma$\,Orionis, purely
photospheric, and $\sigma$\,Ori IRS1, reproduced by a combination of thermal
dust radiation ($T_{\rm dust}=130$ K) and free-free emission.}
\end{figure}

Photo-evaporation of the gas drives a mass-loss rate of $\dot{M}$[M$_\odot$
yr$^{-1}$]$\simeq7.5\times10^{-6}(L_{\rm H}/10^{49})^\frac{1}{2}(d/1 {\rm
pc})^{-1}(r/0.1 {\rm pc})^\frac{3}{2}$ (Bally \& Reipurth 2001), which for
$\sigma$\,Ori IRS1 results in a mass-loss rate of
$\dot{M}\simeq7\times10^{-7}$ M$_\odot$ yr$^{-1}$. If the estimated ion
density of $n\simeq10^6$ cm$^{-3}$ is representative for a uniform gas density
in a region of $r\simeq300$ AU, then the total gas mass is of order $M_{\rm
gas}\sim10^{-3}$ M$_\odot$. The photo-evaporation timescale is thus much
shorter than the age of $\sigma$\,Orionis. Therefore, there must be a
reservoir of neutral material feeding the ionization front. As the radial
density profile is expected to rise towards the centre of IRS1, the gas mass
estimate should be regarded as a lower limit to the total amount of gas
contained within the object.

The H$\alpha$ surface brightness of the ionization front is $I$[erg s$^{-1}$
cm$^{-2}$ arcsec$^{-2}$]$\simeq2.0\times10^{-18}n_{\rm e}^2(r/1 {\rm pc})$
(Spitzer 1978), where we have assumed that the line-of-sight depth of the
region is approximately $r$. We thus obtain $I\simeq3\times10^{-9}$ erg
s$^{-1}$ cm$^{-2}$ arcsec$^{-2}$. For $\sigma$\,Orionis, the flux density
around H$\alpha$ is $F\simeq1.3\times10^{-9}$ erg s$^{-1}$ cm$^{-2}$
nm$^{-1}$. Through a narrow-band H$\alpha$ filter with $\Delta\lambda$ of a
few nm, the (surface) brightness of the ionized region in IRS1 would appear
very similar to that of $\sigma$\,Orionis at a seeing of
$\simeq1^{\prime\prime}$, and at $>2^{\prime\prime}$ distance it would be
easily resolved.

\subsection{Dust properties}

The IRAS position is $\sim5^{\prime\prime}$ North and $\sim12^{\prime\prime}$
East from $\sigma$\,Orionis, with an error ellipse of
$18^{\prime\prime}\times5^{\prime\prime}$ under an angle of $87^\circ$ from
North to East. This is more consistent with the location of $\sigma$\,Ori IRS1
than $\sigma$\,Orionis. Indeed, whilst the spectral energy distribution of
$\sigma$\,Orionis follows the Rayleigh-Jeans tail of its photospheric
continuum emission, the continuum of the mid-IR spectrum and photometry of
$\sigma$\,Ori IRS1 can be modelled by a Planck curve with a (dust) temperature
of $T_{\rm dust}=130\pm10$ K (Fig.\ 3).

The structure in the dust emission feature of $\sigma$\,Ori IRS1 shows clear
evidence for the dust to have been processed compared to interstellar dust.
The latter is dominated by small amorphous silicate grains, with a typical
radius of $a\simeq0.1$ $\mu$m, of which the emission feature peaks sharply at
$\lambda=9.8$ $\mu$m and has a gradual downward slope towards $\lambda\sim12$
$\mu$m. The observed feature in the spectrum of $\sigma$\,Ori IRS1 is much
broader, and peaks at a significantly longer wavelength. Bouwman et al.\
(2001) show that the overall breath of the feature may be reproduced by larger
grains, with a radius $a\simeq1$ $\mu$m, that the peak at $\lambda\simeq11.2$
$\mu$m indicates the presence of Mg-rich crystalline silicate (forsterite),
and that the shoulder at $\lambda\simeq8.6$ $\mu$m can be explained by silica
(SiO$_2$). Such a mixture is often seen in environments in which the dust is
expected to have been processed by co-agulation, leading to grain growth, and
thermal annealing, producing crystalline grains and silica. Indeed, the
spectrum of $\sigma$\,Ori IRS1 resembles that of $\beta$ Pictoris and comet
Hale-Bopp (Bouwman et al.\ 2001).

An estimate of the total dust mass may be obtained by assuming that the dust
emission is optically thin at mid-IR wavelengths. For an ensemble of $N$ dust
grains of radius $a$, at a distance $d_\oplus$ from Earth, the observed flux
density is $F_\lambda=N(a/d_\oplus)^2\pi B_\lambda$. The dust emission can be
described by a Planck curve with a single dust temperature $T_{\rm dust}=130$
K, and it therefore does not matter at which wavelength we evaluate the flux
density. At $\lambda=25$ $\mu$m, $F_\lambda=15$ Jy, and with $a=1$ $\mu$m and
$d_\oplus=352$ pc we obtain $N\simeq2\times10^{37}$ dust grains. At a typical
density of $\rho=2.5$ g cm$^{-3}$, this corresponds to a total dust mass of
(only) $M_{\rm dust}\sim10^{-7}$ M$_\odot$. We note, however, that much more
mass may be hidden in larger grains, rocks and planetesimals without being
observable. From the estimated gas mass of $M_{\rm gas}>10^{-3}$ M$_\odot$,
for a reasonable gas-to-dust mass ratio of order $10^2$, a dust mass of
$M_{\rm dust}>10^{-5}$ M$_\odot$ would be expected --- which indeed suggests
that most of the dust is locked up in larger pockets than the 1 $\mu$m grains
that dominate the mid-IR radiation.

A certain fraction of the light from $\sigma$\,Orionis AB within a solid angle
$\omega$, subtended by IRS1 as seen from $\sigma$\,Orionis, is absorbed by the
dust and re-radiated isotropically: $N4\pi a^2\sigma T_{\rm
dust}^4<(\omega/4\pi)L_\star$. With $L_\star\simeq100,000$ L$_\odot$, we
obtain a solid angle of $\omega>0.0013$ sterad. With a radius $R\simeq300$ AU,
this corresponds to an upper limit to the distance from $\sigma$\,Orionis of
$d<15,000$ AU. This agrees with the morphology of IRS1, the location of the
radio source, and the estimates for the electron density in the ionization
front, which all suggest that we have a (roughly) side-view of the
IRS1-$\sigma$\,Orionis couple, with a separation close to the projected
distance of $d\simeq1200$ AU.

\section{The nature of $\sigma$\,Ori IRS1}

The dust cloud $\sigma$\,Ori IRS1 measures $\sim10^3$ AU across and contains
$>10^{-3}$ M$_\odot$ of gas and dust, the latter of which shows signs of
co-agulation. This suggests that IRS1 may be a proto-planetary disk. Indeed,
circumstellar disks have been detected around several T Tauri stars in the
$\sigma$\,Orionis cluster (Oliveira et al.\ 2003). As expected, the close
proximity of IRS1 to $\sigma$\,Orionis gives rise to an ionization front at
the side of IRS1 facing the O star, and it thus resembles the
photo-evaporating proto-planetary disks seen in H\,{\sc ii} regions such as
the Orion Nebula, otherwise known as ``proplyds'' (O'Dell et al.\ 1993). The
rapid photo-destruction timescale is a problem though, unless the total mass
contained within the disk of $\sigma$\,Ori IRS1 amounts to at least one solar
mass or the object has been (much) further away during most of its past
lifetime.

However, there is no evidence yet that $\sigma$\,Ori IRS1 hosts a central
star. Hence the possibility remains that IRS1 is instead a dense knot of
interstellar material --- albeit processed --- perhaps similar to the starless
small clouds discovered recently in the Carina Nebula, a 3 Myr old massive
star forming region (Smith, Bally \& Morse 2003). The proximity of IRS1 to
$\sigma$\,Orionis may well be only temporary, alleviating the problem of the
rapid photo-destruction. This would imply that such clouds must be rather
common around $\sigma$\,Orionis, something which can be tested by means of a
deep survey of the entire $\sigma$\,Orionis cluster at mid/far-IR or (sub)mm
wavelengths.

\begin{acknowledgements}
We thank Dr.\ Michael Sterzik for support at the telescope, Prof.\ Rens Waters
for discussion on the dust species, and Prof.\ O'Dell for his constructive
referee report. The IRAS data base server of the Space Research Organisation
of the Netherlands (SRON) and the Dutch Experise Centre for Astronomical Data
Processing is funded by the Netherlands Organisation for Scientific Research
(NWO). The IRAS data base server project was also partly funded through the
Air Force Office of Scientific Research, grants AFOSR 86-0140 and
AFOSR 89-0320. JMO acknowledges support of the UK Particle Physics and
Astronomy Research Council.
\end{acknowledgements}


\begin{thebibliography}{}
\bibitem[2001]{BallyReipurth2001}
Bally J., Reipurth B., 2001, ApJ 546, 299
\bibitem[2001]{BouwmanEtal2001}
Bouwman J., Meeus G., de Koter A., et al., 2001, A\&A 375, 950
\bibitem[2000]{BrandnerEtal2000}
Brandner W., Grebel E.K., Chu Y.-H., et al., 2000, AJ 119, 292
\bibitem[1994]{BrownDegeusDezeeuw1994}
Brown A.G.A., de Geus E.J., de Zeeuw P.T., 1994, A\&A 289, 101
\bibitem[1991]{ChlebowskiGarmany1991}
Chlebowski T., Garmany C.D., 1991, ApJ 368, 241
\bibitem[1990]{Drake1990}
Drake S.A., 1990, AJ 100, 572
\bibitem[2001]{HaischLadaLada2001}
Haisch K.E., Lada E.A., Lada C.J., 2001, ApJ 553, L153
\bibitem[1997]{Heintz1997}
Heintz W.D., 1997, ApJS 111, 335
\bibitem[1995]{HofmannSeggewissWeigelt1995}
Hofmann K.-H., Seggewiss W., Weigelt G., 1995, A\&A 300, 403
\bibitem[1974]{Lucy1974}
Lucy L.B., 1974, AJ 79,745
\bibitem[1994]{MccaughreanStauffer1994}
McCaughrean M.J., Stauffer J.R., 1994, AJ 108, 1382
\bibitem[1993]{OdellWenHu1993}
O'Dell C.R. Wen Z., Hu X., 1993, ApJ 410, 696
\bibitem[2002]{OliveiraEtal2002}
Oliveira J.M., Jeffries R.D., Kenyon M.J., Thompson S.A., Naylor T., 2002,
A\&A 382, L22
\bibitem[2003]{OliveiraEtal2003}
Oliveira J.M., Jeffries R.D., van Loon J.Th., Kenyon M.J., 2003, in: Open
Issues of Local Star Formation and Early Stellar Evolution, eds.\ J.\
Gregorio-Hetem \& J.\ L\'{e}pine
\bibitem[1974]{Osterbrock1974}
Osterbrock D.E., 1974, ``Astrophysics of Gaseous Nebulae'', W.H.\ Freeman and
Company, San Francisco, p.79
\bibitem[1975]{PeimbertEtal1975}
Peimbert M., Rayo J.F., Torres-Peimbert S., 1975, Rev.\ Mex.\ Astron.\
Astrofis.\ 1, 289
\bibitem[2003]{SmithBallyMorse2003}
Smith N., Bally J., Morse J.A., 2003, ApJ 587, L105
\bibitem[1978]{Spitzer1978}
Spitzer L., 1978, ``Physical Processes in the InterMedium'', Wiley, New York
\bibitem[1997]{WalterWolkSherry1997}
Walter F.M., Wolk S.J., Sherry W., 1997, in: The $10^{\rm th}$ Cambridge
Workshop on Cool Stars, Stellar Systems and the Sun, eds.\ R.A.\ Donahue \&
J.A.\ Bookbinder, ASP Conf.Ser.\ 154, p1793
\bibitem[2000]{ZapateroosorioEtal2000}
Zapatero Osorio M.R., B\'{e}jar V.J.S., Mart\'{\i}n E.L., et al., 2000,
Science 290, 103
\end{thebibliography}
\end{document}